# Tunable electronic properties and electric-field-induced phase transition in phosphorene/graphene heterostructures

Maryam Mahdavifar, Sima Shekarforoush, and Farhad Khoeini*

Department of Physics, University of Zanjan, P. O. Box 45195-313, Zanjan, Iran

**Abstract:** The shortcomings of mono-component systems, e.g., the gapless nature of graphene, the lack of air-stability in phosphorene, etc. have drawn great attention toward stacked materials expected to show interesting electronic and optical properties. Using the tight-binding approach and the Green's function method, we investigate the electronic properties of armchair-edged lateral phosphorene/graphene heterostructures, which are either semiconductor/semiconductor or semiconductor/metal heterostructures, depending on the width of graphene ribbon. It is found that the system is narrow-gapped, and the bandgap can be modulated by tuning the size of the domains. Besides, the analysis of the bandgap variation against the width of the component phosphorene ribbon indicates that, in semiconductor/metal heterostructure, phosphorene ribbon does not induce any electronic state near the Fermi level, suggesting that the suppressed electron transport should be attributed to the hole transfer across the interface. Furthermore, we show that the transverse electric field can significantly diversify the electronic behavior of the heterostructure, i.e., the heterostructure undergoes the semiconductor-metal phase transition. Moreover, tuning the transverse electric field yields an intriguing possibility that the system can undergo a topological phase transition from a band insulator to a topological insulator.

## 1. Introduction

The structural properties of two dimensional (2D) materials, such as atomic-level thickness, surface inertness, single layer or few-layer arrangement, and low dimensionality allow forming the heterostructures [1, 2]. A heterostructure consists of two or more types of materials with various functionalities, which may be stacked either vertically (on top of each other) or laterally (in-plane) [3]. Compared to the vertical heterostructures, the lateral heterostructures have smaller interface area and simpler band alignment [4, 5]. Besides, unlike the vertical ones, which are generally called van der Waals heterostructures, distinct domains of a lateral heterostructure are usually stacked by covalent bonds, ensuring the epitaxial quality of boundaries by one and two-step CVD processes, as recently carried out lateral $WSe_2/MoS_2$ heterostructure with an atomically sharp interface through a two-step epitaxial growth [5-7]. Thereby, lateral heterostructures should be rather promising for many practical applications.

Shortcomings of mono-component systems have drawn great attention toward stacked materials leading to interesting electronic and optical properties [8-15]. The mobile electrons in graphene behave as massless Dirac fermions making it an important material in fundamental physics [16-19]. However, due to the gapless nature of graphene, there is no possibility for electrically manipulating its band structure [20, 21]. Unlike graphene, phosphorene is an intrinsic p-type semiconductor material with a direct bandgap of about 1.5 eV [22-26]. Phosphorene also reveals unique structural properties such as high mechanical flexibility, strong structural anisotropy, and biocompatibility [27, 28]. Moreover, phosphorene has a high carrier mobility of about 1000 $cm^2$/V.s. and a moderate on/off current ratio ($10^4$), which makes it attractive for field-effect transistors [29, 30]. Despite the excellent electronic properties of phosphorene, the lack of air-stability in phosphorene is a key challenge [31, 32]. To overcome the issues and challenges associated with the shortcomings of 2D structures and their transformations, one may resort to stacked materials [5].

Ever since, a large number of stacked materials have been considered, which have shown exotic properties. For example, Padilha et al. calculated the electronic band structure of vertical phosphorene/graphene heterostructure and showed that it is possible to control the Schottky barrier using a perpendicular electric field [33]. To protect phosphorene from structural and chemical degradation, Cai et al. have proposed a graphene or a hexagonal boron nitride layer vertically stacked to phosphorene. They showed that the interesting properties of phosphorene are maintained in such heterostructures [34]. The electronic properties of the lateral heterostructure comprised of black and blue phosphorene have been investigated by Li et al., showing that using size effect and strain engineering, it is easy to tune bandgaps, carrier effective masses, and band alignments [35].

In this work, motivated by the high potential of the heterostructures for innovative technological applications, we investigate the electronic properties of lateral phosphorene/graphene heterostructures using the tight-binding model and Green's function method. The system is comprised of an armchair graphene nanoribbon (AGNR) and armchair phosphorene nanoribbon (APNR). The APNR exhibits the semiconducting behavior [22], while, the AGNR can be either gapless (for the width of $w = 3n + 2$, where $n$ is an integer) or gapped (for the width of $w = 3n, 3n + 1$, where $n$ is an integer) [36]. Therefore, the system is either semiconductor/semiconductor or semiconductor/metal heterostructure, depending on the AGNR width. However, we show that the system is narrow-gapped such that one can tune the bandgap by tuning the size of the domains. This result is in a good agreement with the results derived from density functional theory calculations reported by Tian et al. [37]. They have also shown the tunability of the electronic features by the applied perpendicular electric field. The analysis of the bandgap variation against the width of the component phosphorene ribbon indicates that, in the vicinity of the Fermi level, the component APNR does not induce any electronic state, suggesting that the suppressed electron transport should be attributed to the interface region formed by these dissimilar materials. Indeed, a hole transfer occurs across the interface from APNR into metallic AGNR, which results in a reduction of the electron transfer through the AGNR. It also found that a transverse electric field can significantly diversify the electronic behavior of the heterostructure, i.e., the heterostructure can be either gapped or gapless under the transverse electric field and undergoes the semiconductor-metal phase transition. Moreover, tuning the transverse electric field yields an intriguing possibility that a system can undergo a topological phase transition from a band insulator to a topological insulator.

The rest of the article is organized as follows. Section II gives the design of the model, and describes the method in detail. Section III consists of two parts: first, we investigate the electronic properties of the system, then we study the electronic behavior of the system subjected to the electric field. Finally, we conclude with a summary in section IV.

**2. Model and method**

**2.1. Structural properties of graphene**

Graphene due to the in-plane $sp^2$ hybridized orbitals has a planar honeycomb structure with the carbon-carbon distance of 1.42 Å, consisting of two sublattices A and B [38]. Cutting the graphene sheet along different directions can yield graphene nanoribbons (GNRs) with very different electronic properties depending on their edge configuration and width [39]. For instance, all zigzag-edged ribbons show metallic features, while the armchair-edged ribbons can be either metallic or semiconducting depending on the width [40]. This feature allows GNRs to be helpful in designing heterostructures as functional elements. In this work, the AGNR is chosen to be a component of the system, as shown in figure 1. We use $N_G$, the number of A (B)-site carbon atoms per unit cell (see figure 1), to denote different AGNR widths, and $N_G$-AGNR to determine a typical AGNR width.

**2.2. Structural properties of phosphorene**

Monolayer phosphorene mainly due to the $sp^3$ hybridized orbitals has a puckered honeycomb structure such that each P atom bonds with two P atoms in-plane and one P atom out of plane [24]. The inter-planar P-P distance is 2.259 Å, and the intra-planar P-P distance is 2.220 Å, using PBE [41]. Accordingly, it has a primitive unit cell containing four phosphorus atoms.

Phosphorene has a thickness-dependence bandgap, which remains direct for any number of layers and changes from 1.3 to 2 eV (for 1-5 layers) [26]. However, the bandgap can be controlled by cutting the phosphorene sheet into the ribbons [42]. Here, $N_P$, the number of pairs of phosphorus atoms per the unit-cell (see figure 1), is used to denote the width of the armchair phosphorene nanoribbon. We use $N_P$-APNR to denote a typical APNR with the width of $N_P$.

**2.3. Structural properties of phosphorene/graphene heterostructure**

A phosphorene/graphene heterostructure is made of an APNR and an AGNR laterally connected along the armchair direction, as shown in figure 1. There is a small lattice mismatch of 2.6 percent between the ribbons along with the interface. The P–C distance at the interface has been adopted from the DFT results assumed to be within the interval of C–C and P–P distances [37]. $N$ (=$N_P$+$N_G$) is used to indicate the width of the APNR/AGNR heterostructure.

In our calculation, the system is composed of three regions: the central region with a confined length connected to two semi-infinite electrodes. According to figure 1, the central region is characterized by M unit cells along the x-axis.

**2.4. Tight-binding model**

To calculate the electronic band structure and electronic transport through the heterostructure, we use a low-energy effective Hamiltonian based on the tight-binding model with one π-electron per atom, described as

$$\boldsymbol{H}_{P/G/C} = \sum_i \varepsilon_i c_i^\dagger c_i + \sum_{i \neq j} t_{i,j} c_i^\dagger c_j + eE_y \sum_i d_i c_i^\dagger c_i, \tag{1}$$

where $\boldsymbol{H}_{P/G/C}$ denotes the Hamiltonian of the APNR, AGNR, and the coupling between the ribbons, respectively. Moreover, $c_i^\dagger$ and $c_i$ denote the creation (annihilation) operator of electrons at site $i$, $\varepsilon_i$ is the on-site energy at site $i$, and $t_{i,j}$ is the hopping energy between sites $i$ and $j$. The third term describes the transverse electric field effect, where $E_y$ denotes the electric field, which is applied to the system, as shown in figure 1, and $d_i$ is the vertical distance of atom $i$ from the bottom edge of the system [43].

In this work, we use a simplified tight-binding model that includes five hopping parameters for $3p_z$ orbitals of phosphorus atoms (as depicted in figure 1) [44], and a model in which the hopping energy between carbon atoms is taken up to third-nearest-neighbor [45]. The hopping parameter of the interface region, including the coupling of C and P atoms, is obtained by the method introduced in Ref. [46]. We extracted the on-site energies by fitting the band structure of the heterostructure to DFT calculations in which the carbon and phosphorus on-site energies were only the fitting parameters. The on-site and hopping energies of the system are reported in Table 1.

For an infinite ribbon, the electron wave vector along the x-axis ($k$) is a good quantum number. Therefore, by applying the Bloch's theorem, the $k$-dependent Hamiltonian of the system can be written as

$$\boldsymbol{H}_k = \boldsymbol{H}_{00} + \boldsymbol{H}_{-10}\exp(ika) + \boldsymbol{H}_{-10}^\dagger \exp(-ika), \qquad (2)$$

where $\boldsymbol{H}_{00}$ is the unit cell Hamiltonian, and $\boldsymbol{H}_{-10}$ is the hopping Hamiltonian between the neighboring cells. We calculate the band structure of the heterostructure by diagonalizing the Hamiltonian (2). The unit cell Hamiltonian matrix is defined as

$$\boldsymbol{H}_{00} = \begin{pmatrix} \boldsymbol{H}_P & \boldsymbol{H}_C \\ \boldsymbol{H}_C^\dagger & \boldsymbol{H}_G \end{pmatrix}. \qquad (3)$$

**2.5. Recursive Green's function method**

We utilize Green's function method with a quick iterative scheme to obtain the density of states (DOS) and the electron transmission probability. The effective Green's function is given by

$$G_D(E) = ((E + i\eta)\mathbf{I} - \mathbf{H}_D - \mathbf{\Sigma}_L(E) - \mathbf{\Sigma}_R(E))^{-1}, \tag{4}$$

where η is an infinitesimal positive real constant, $\mathbf{I}$ is the identity matrix, and $\mathbf{H}_D$ is the Hamiltonian of the device. Besides, $\mathbf{\Sigma}_{L(R)}$ is the left (right) self-energy operator described as

$$\mathbf{\Sigma}_{L(R)} = \mathbf{H}^\dagger_{LD(RD)} \mathbf{g}_{L(R)} \mathbf{H}_{LD(RD)}, \tag{5}$$

where $\mathbf{H}_{LD(RD)}$ is the coupling between the device and the left (right) lead [46-48]. It should be noticed that the coupling between the device and the leads is assumed to be perfect. Besides, $\mathbf{g}_{L(R)}$ is the surface Green's function of the left (right) lead, which has been described by Sancho et al. [49, 50] as

$$\mathbf{g}_L = [(E + i\eta)\mathbf{I} - \mathbf{H}_{00} - \mathbf{H}^\dagger_{-10}\widetilde{\mathbf{P}})]^{-1}, \tag{6}$$

$$\mathbf{g}_R = [(E + i\eta)\mathbf{I} - \mathbf{H}_{00} - \mathbf{H}_{-10}\mathbf{P})]^{-1}, \tag{7}$$

where $\mathbf{I}$ is a unit matrix. Moreover, $\mathbf{P}$ and $\widetilde{\mathbf{P}}$ are the transfer matrices, which are calculated from the Hamiltonian matrix elements through an iterative procedure as

$$\mathbf{P} = \mathbf{p}_0 + \widetilde{\mathbf{p}}_0 \mathbf{p}_1 + \widetilde{\mathbf{p}}_0 \widetilde{\mathbf{p}}_1 \mathbf{p}_2 + \cdots + \widetilde{\mathbf{p}}_0 \widetilde{\mathbf{p}}_1 \widetilde{\mathbf{p}}_2 \ldots \mathbf{p}_n, \tag{8}$$

$$\widetilde{\mathbf{P}} = \widetilde{\mathbf{p}}_0 + \mathbf{p}_0 \widetilde{\mathbf{p}}_1 + \mathbf{p}_0 \mathbf{p}_1 \widetilde{\mathbf{p}}_2 + \cdots + \mathbf{p}_0 \mathbf{p}_1 \mathbf{p}_2 \ldots \widetilde{\mathbf{p}}_n, \tag{9}$$

where

$$\mathbf{p}_0 = [(E + i\eta)\mathbf{I} - \mathbf{H}_{00}]^{-1} \mathbf{H}^\dagger_{-10}, \tag{10}$$

$$\widetilde{\mathbf{p}}_0 = [(E + i\eta)\mathbf{I} - \mathbf{H}_{00}]^{-1} \mathbf{H}_{-10}, \tag{11}$$

$$\mathbf{p}_i = (\mathbf{I} - \mathbf{p}_{i-1}\widetilde{\mathbf{p}}_{i-1} - \widetilde{\mathbf{p}}_{i-1}\mathbf{p}_{i-1})^{-1} \mathbf{p}^2_{i-1}, \tag{12}$$

$$\widetilde{\mathbf{p}}_i = (\mathbf{I} - \mathbf{p}_{i-1}\widetilde{\mathbf{p}}_{i-1} - \widetilde{\mathbf{p}}_{i-1}\mathbf{p}_{i-1})^{-1} \widetilde{\mathbf{p}}^2_{i-1}. \tag{13}$$

Transmission probabilities can be calculated from the effective Green's function and the coupling between the device and the leads in the coherent regime

$$T(E) = \text{trace}(\pmb{\Gamma}_\text{L}(E)\pmb{G}_\text{D}(E)\pmb{\Gamma}_\text{R}(E)\pmb{G}_\text{D}(E)^\dagger), \quad (14)$$

where

$$\pmb{\Gamma}_\text{L(R)}(E) = \text{i}\big[\pmb{\Sigma}_\text{L(R)}(E) - (\pmb{\Sigma}_\text{L(R)}(E))^\dagger\big]. \quad (15)$$

Finally, the conductance is calculated by the Landauer formula, described as

$$G = 2\frac{\text{e}^2}{\text{h}}T. \quad (16)$$

and the DOS at energy $E$ is given by

$$DOS(E) = -\frac{1}{\pi}\text{Im}(\text{trace}(\pmb{G}_\text{D}(E))). \quad (17)$$

### 3. Numerical results and discussion

#### 3.1. Electronic properties

Here, we numerically study the electronic properties of the armchair-edged phosphorene/graphene heterostructure based on the method addressed in Section 2. The right panel of figure 1 shows the band structure of APNR/AGNR system with the width of $N = 71$ ($w \sim 10nm$), comprised of the two ribbons with the widths of $N_\text{P} = 30$ ($w \sim 5nm$) and $N_\text{G} = 41$ ($w \sim 5nm$), which represents a small direct gap of 0.08 eV. The energy spectra of the isolated ribbons are shown in the left panels of the figure. The conductance and the DOS spectra of the heterostructure and the isolated component ribbons depicted in figure 2, are in agreement with the corresponding energy spectra. The energy gap of APNR and AGNR as functions of the ribbon width are shown in figure 3 (a) (yellow circles) and 3 (b) (gray circles), respectively. The APNRs exhibit the semiconducting feature with a bandgap greater than 1.5 eV, while the AGNRs are either metallic or semiconducting. Since 41-AGNR has the metallic feature, the energy gap opened in the heterostructure seems to be the nontrivial gap. However, the zero conductance of the semiconductor/metal heterostructure at low energy indicates that the electronic transport is relatively suppressed through such a system.

In figure 2, another point to note is the staircase-like conductance of the heterostructure, arising from the negligible mismatch of the component ribbons and the lack of dangling bonds. A small lattice mismatch between two in-plane stacked materials can result in a clear boundary in the interface, such as the growth of lateral $WSe_2/MoS_2$ heterostructure or lateral graphene/boron nitride heterostructure, which have shown an atomically sharp interface despite the small lattice mismatch between the stacked materials [7, 9]. If there exist imperfections at the interface region, it may induce a deviation of conductance from the staircase-like curve [51].

Figure 3(a)-(b) presents the bandgap of $N_P$-APNR/41-AGNR heterostructure against $N_P$ and 30-APNR/$N_G$-AGNR heterostructure against $N_G$ (purple circles), respectively. As shown in figure 3(a), the bandgap of the heterostructure remains invariant (∼0.08 eV) against the various widths of the component APNR. It demonstrates that the electronic property of the heterostructure near the Fermi level is independent of the electron transport through the APNR, suggesting that the APNR does not induce any electronic state to the charge transmission near the Fermi energy. Thereby, the suppressed conductance of the semiconductor/metal heterostructure with metallic component AGNR, near the Fermi energy, should be attributed to the interface region. We also inspected the band structure of semiconductor/semiconductor heterostructure, which implied a decrease in the bandgap by increasing the width of the component phosphorene ribbon for narrow graphene ribbons ($N_G < 11$). By the way, figure 3(b) shows that the energy gap of the heterostructure mostly depends on the electronic behavior of the AGNR. Therefore, the bandgap can be modulated by tuning the width of the regions. According to the figure, a small bandgap opens in the semiconductor/metal heterostructures

($N_G = 3n + 2$). This is due to a charge carrier (hole) transfer across the interface from APNR into AGNR arising from the coupling between the metal and the intrinsic p-type semiconductor materials, which results in a reduction of the electron transmission through the heterostructure [33, 52].

However, the results show that the heterostructure is a narrow-gapped system against any width of AGNR. The bandgap of semiconductor/semiconductor heterostructure is smaller (greater) than that of the corresponding isolated AGNR with the width of $N_G = 3n$ ($3n + 1$). For instance, here, we analyze the electronic behavior of 30-APNR/39-AGNR ($E_g = 0.13$ eV) and 30-APNR/43-AGNR ($E_g = 0.27$ eV). We found that the bandgap of each heterostructure corresponds to a typical isolated AGNR; near the Fermi level 30-APNR/39-AGNR behaves like an isolated 54-AGNR, and 30-APNR/43-AGNR behaves like an isolated 34-AGNR. The local density of states (LDOS) of conduction band minimum (CBM) and valence band maximum (VBM) of the heterostructures compared to the corresponding isolated AGNRs are shown in figure 4(a)-(d). The atomic sites of each ribbon are labeled atom by atom from the lower edge to the upper edge (see figure 1). Depicted in figure 4(a)-(b), the LDOS distribution of CBM and VBM of the component 39-AGNR (red circles) mostly match with those of 54-AGNR (blue stars) rather than 39-AGNR (green triangular), respectively. Namely, the electronic states of CBM and VBM of 30-APNR/39-AGNR mostly behave like 54-AGNR, giving rise to a similar electronic bandgap. The same reasoning is applied to 30-APNR/43-AGNR. As shown in figure 4(c)-(d), also here, the LDOS distribution of CBM and VBM of the component 43-AGNR (red circles) mostly matches with those of the isolated 34-AGNR (blue stars) rather than the isolated 43-AGNR (green triangular), respectively. Thereby, the heterostructure reflects the electronic properties of 34-AGNR rather than 43-AGNR. We mentioned earlier that the APNR does not induce any electronic state to the charge transfer near the Fermi level for wide AGNRs, which is also implied by the LDOS distribution in the heterostructures. But, the electronic behavior is mainly affected through the interface region formed by these two ribbons.

### 3.2. Bandgap modification by electric field

The electric field has been identified as one of the strategies to highly tune the bandgap of 2D structures and their derivatives [43]. In the following, the armchair-edged phosphorene/graphene heterostructure is assumed to be present under the transverse electric field. The analysis of the electronic bandgap of the system and its isolated component ribbons under the transverse electric field will be done by inspection of figure 5. Figure 5(a)-(c) shows the bandgap of 30-APNR, 41-AGNR, and 30-APNR/41-AGNR heterostructure as functions of a transverse electric field, respectively. The bandgap of the AGNR and APNR slightly increases and decreases, respectively, with increasing the transverse electric field. However, the transverse electric field can significantly diversify the electronic properties of the heterostructure. The variation of the bandgap with the applied transverse electric field yields the critical electric fields at $E_{yc} = 0.013$ V/Å and $-0.009$ V/Å. Moreover, according to the figure, at $E_y = 0.015$ V/Å and $E_{yc} = -0.011$ V/Å (and higher values of them, regardless the sign), the system is gapless. Therefore, the APNR/AGNR heterostructure can be gapped or gapless under the transverse electric field and undergoes the semiconductor-metal phase transition.

We show the band structure of the heterostructure in the absence of the transverse electric field and the presence of $E_y = 0.016$ V/Å in figure 6(a)-(b), respectively. The red and blue lines show the edge states. When the transverse electric field is set to zero, the system is a band insulator. In the presence of the transverse electric field of $0.016$ V/Å, the edge states are gapless while the bulk states are gapped, which shows that the system is a topological insulator. Consequently, the electronic band structure of the APNR/AGNR heterostructure is easily tuned by a transverse electric field, which implies an intriguing possibility that the system can undergo a topological phase transition.

## 4. Summary and conclusion remarks

We investigated the electronic behavior of the armchair-edged phosphorene/graphene heterostructures, which are either semiconductor/semiconductor or semiconductor/metal heterostructures, depending on the width of AGNR. In summary, it was found that the heterostructure is narrow-gapped and the bandgap can be modulated by tuning the size of the widths. Moreover, the bandgap variation against the width of the component APNR demonstrated that, in semiconductor/metal heterostructure, phosphorene ribbon does not induce any electronic state to the electron transfer through the heterostructure, near the Fermi level. This suggests that the suppressed electron transport should be attributed to the interface region, which is associated with the hole transfer across the semiconductor/metal interface, from intrinsic p-type APNR into metallic AGNR. We found that the bandgap of semiconductor/semiconductor heterostructures are smaller (greater) than that of the corresponding isolated AGNRs with the width of $N_G = 3n\ (3n + 1)$. This was reasoned by the LDOS distribution of CBM and VBM of the system, which showed that each heterostructure reflects the electronic properties of a typical isolated AGNR near the Fermi level.

We also studied the bandgap modification of the heterostructures by a transverse electric field. We showed that the transverse electric field could significantly diversify the electronic behavior of the heterostructures. Namely, the semiconductor-metal phase transition is possible by tuning the transverse electric field. Moreover, tuning the transverse electric field yields an intriguing possibility that the system can undergo a topological phase transition from a band insulator to a topological insulator.

*Corresponding Author's Email: khoeini@znu.ac.ir

**References**

[1] Miró P, Audiffred M, Heine T. An atlas of two-dimensional materials. Chemical Society Reviews. 2014;43(18):6537-54.

[2] Pant A, Mutlu Z, Wickramaratne D, Cai H, Lake RK, Ozkan C, et al. Fundamentals of lateral and vertical heterojunctions of atomically thin materials. Nanoscale. 2016;8(7):3870-87.

[3] Wang J, Li Z, Chen H, Deng G, Niu X. Recent advances in 2D lateral heterostructures. Nano-Micro Letters. 2019;11(1):48.

[4] Zhang J, Xie W, Zhao J, Zhang S. Band alignment of two-dimensional lateral heterostructures. 2D Materials. 2016;4(1):015038.

[5] Zhao J, Cheng K, Han N, Zhang J. Growth control, interface behavior, band alignment, and potential device applications of 2D lateral heterostructures. Wiley Interdisciplinary Reviews: Computational Molecular Science. 2018;8(2):e1353.

[6] Duesberg GS. Heterojunctions in 2D semiconductors: A perfect match. Nature materials. 2014;13(12):1075-6.

[7] Li M-Y, Shi Y, Cheng C-C, Lu L-S, Lin Y-C, Tang H-L, et al. Epitaxial growth of a monolayer WSe2-MoS2 lateral pn junction with an atomically sharp interface. Science. 2015;349(6247):524-8.

[8] Britnell L, Gorbachev R, Jalil R, Belle B, Schedin F, Mishchenko A, et al. Field-effect tunneling transistor based on vertical graphene heterostructures. Science. 2012;335(6071):947-50.

[9] Levendorf MP, Kim C-J, Brown L, Huang PY, Havener RW, Muller DA, et al. Graphene and boron nitride lateral heterostructures for atomically thin circuitry. Nature. 2012;488(7413):627-32.

[10] Bertolazzi S, Krasnozhon D, Kis A. Nonvolatile memory cells based on MoS2/graphene heterostructures. ACS nano. 2013;7(4):3246-52.

[11] Geim AK, Grigorieva IV. Van der Waals heterostructures. Nature. 2013;499(7459):419-25.

[12] Zhang W, Chuu C-P, Huang J-K, Chen C-H, Tsai M-L, Chang Y-H, et al. Ultrahigh-gain photodetectors based on atomically thin graphene-MoS 2 heterostructures. Scientific reports. 2014;4:3826.

[13] Barrios-Vargas JE, Mortazavi B, Cummings AW, Martinez-Gordillo R, Pruneda M, Colombo L, et al. Electrical and thermal transport in coplanar polycrystalline graphene–hbn heterostructures. Nano letters. 2017;17(3):1660-4.

[14] Ju L, Dai Y, Wei W, Li M, Huang B. DFT investigation on two-dimensional GeS/WS2 van der Waals heterostructure for direct Z-scheme photocatalytic overall water splitting. Applied Surface Science. 2018;434:365-74.

[15] Bafekry A, Akgenc B, Shayesteh SF, Mortazavi B. Tunable electronic and magnetic properties of graphene/carbon-nitride van der Waals heterostructures. Applied Surface Science. 2020;505:144450.

[16] Novoselov KS, Geim AK, Morozov SV, Jiang D, Katsnelson MI, Grigorieva I, et al. Two-dimensional gas of massless Dirac fermions in graphene. nature. 2005;438(7065):197-200.

[17] Nomura K, MacDonald A. Quantum transport of massless Dirac fermions. Physical review letters. 2007;98(7):076602.

[18]     Khoeini F. Analytical study of electronic quantum transport in carbon-based nanomaterials. Diamond and related materials. 2014;47:7-14.

[19]     Yousefi F, Khoeini F, Rajabpour A. Thermal rectification and interfacial thermal resistance in hybrid pillared-graphene and graphene: a molecular dynamics and continuum approach. Nanotechnology. 2020;31(28):285707.

[20]     Katsnelson M, Novoselov K, Geim A. Chiral tunnelling and the Klein paradox in graphene. Nature physics. 2006;2(9):620-5.

[21]     Ezawa M. A topological insulator and helical zero mode in silicene under an inhomogeneous electric field. New Journal of Physics. 2012;14(3):033003.

[22]     Liu H, Neal AT, Zhu Z, Luo Z, Xu X, Tománek D, et al. Phosphorene: an unexplored 2D semiconductor with a high hole mobility. ACS nano. 2014;8(4):4033-41.

[23]     Tran V, Soklaski R, Liang Y, Yang L. Layer-controlled band gap and anisotropic excitons in few-layer black phosphorus. Physical Review B. 2014;89(23):235319.

[24]     Khoeini F, Nazari M, Shekarforoush S, Mahdavifar M. Electromechanical and magnetic response in zigzag phosphorene nanoribbons. Physica E: Low-dimensional Systems and Nanostructures. 2020:114200.

[25]     Fei R, Yang L. Strain-engineering the anisotropic electrical conductance of few-layer black phosphorus. Nano letters. 2014;14(5):2884-9.

[26]     Guo H, Lu N, Dai J, Wu X, Zeng XC. Phosphorene nanoribbons, phosphorus nanotubes, and van der Waals multilayers. The Journal of Physical Chemistry C. 2014;118(25):14051-9.

[27]     Wei Q, Peng X. Superior mechanical flexibility of phosphorene and few-layer black phosphorus. Applied Physics Letters. 2014;104(25):251915.

[28]     Pumera M. Phosphorene and black phosphorus for sensing and biosensing. TrAC Trends in Analytical Chemistry. 2017;93:1-6.

[29]     Li L, Yu Y, Ye GJ, Ge Q, Ou X, Wu H, et al. Black phosphorus field-effect transistors. Nature nanotechnology. 2014;9(5):372.

[30]     Shekarforoush S, Khoeini F, Shiri D. Optical transitions and localized edge states in skewed zigzag phosphorene nanoribbons. Materials Express. 2018;8(6):489-99.

[31]     Wood JD, Wells SA, Jariwala D, Chen K-S, Cho E, Sangwan VK, et al. Effective passivation of exfoliated black phosphorus transistors against ambient degradation. Nano letters. 2014;14(12):6964-70.

[32]     Favron A, Gaufrès E, Fossard F, Phaneuf-L'Heureux A-L, Tang NY, Lévesque PL, et al. Photooxidation and quantum confinement effects in exfoliated black phosphorus. Nature materials. 2015;14(8):826-32.

[33]     Padilha J, Fazzio A, da Silva AJ. van der Waals heterostructure of phosphorene and graphene: tuning the Schottky barrier and doping by electrostatic gating. Physical review letters. 2015;114(6):066803.

[34]     Cai Y, Zhang G, Zhang Y-W. Electronic properties of phosphorene/graphene and phosphorene/hexagonal boron nitride heterostructures. The Journal of Physical Chemistry C. 2015;119(24):13929-36.

[35]     Li Y, Ma F. Size and strain tunable band alignment of black–blue phosphorene lateral heterostructures. Physical Chemistry Chemical Physics. 2017;19(19):12466-72.

[36]     Mahdavifar M, Khoeini F. Topological and transport properties of graphene-based nanojunctions subjected to a magnetic field. Nanotechnology. 2019;31(2):025701.

[37]     Tian X, Liu L, Du Y, Gu J, Xu J-b, Yakobson BI. Variable electronic properties of lateral phosphorene–graphene heterostructures. Physical Chemistry Chemical Physics. 2015;17(47):31685-92.

[38]     Castro Neto AH, Guinea F, Peres NMR, Novoselov KS, Geim AK. The electronic properties of graphene. RvMP. 2009;81(1):109-62.

[39]     Han MY, Özyilmaz B, Zhang Y, Kim P. Energy band-gap engineering of graphene nanoribbons. Physical review letters. 2007;98(20):206805.

[40]     Brey L, Fertig H. Electronic states of graphene nanoribbons studied with the Dirac equation. Physical Review B. 2006;73(23):235411.

[41]     Sa B, Li Y-L, Qi J, Ahuja R, Sun Z. Strain engineering for phosphorene: the potential application as a photocatalyst. The Journal of Physical Chemistry C. 2014;118(46):26560-8.

[42]     Watts MC, Picco L, Russell-Pavier FS, Cullen PL, Miller TS, Bartuś SP, et al. Production of phosphorene nanoribbons. Nature. 2019;568(7751):216-20.

[43]     Mahdavifar M, Khoeini F. Highly tunable charge and spin transport in silicene junctions: phase transitions and half-metallic states. Nanotechnology. 2018;29(32):325203.

[44]     Rudenko AN, Katsnelson MI. Quasiparticle band structure and tight-binding model for single- and bilayer black phosphorus. Physical Review B. 2014;89(20):201408.

[45]     Hancock Y, Uppstu A, Saloriutta K, Harju A, Puska MJ. Generalized tight-binding transport model for graphene nanoribbon-based systems. Physical Review B. 2010;81(24):245402.

[46]     Qasemnazhand M, Khoeini F, Shekarforoush S. Electronic transport properties in the stable phase of a cumulene/B 7/cumulene molecular bridge investigated using density functional theory and a tight-binding method. New Journal of Chemistry. 2019;43(42):16515-23.

[47]     Khoeini F, Khoeini F, Shokri A. Peculiar transport properties in Z-shaped graphene nanoribbons: A nanoscale NOR gate. Thin solid films. 2013;548:443-8.

[48]     Zhang Z, Xie Y, Peng Q, Chen Y. A theoretical prediction of super high-performance thermoelectric materials based on MoS 2/WS 2 hybrid nanoribbons. Scientific reports. 2016;6:21639.

[49]     Sancho ML, Sancho JL, Rubio J. Quick iterative scheme for the calculation of transfer matrices: application to Mo (100). Journal of Physics F: Metal Physics. 1984;14(5):1205.

[50]     Khoeini F. Combined effect of oriented strain and external magnetic field on electrical properties of superlattice-graphene nanoribbons. Journal of Physics D: Applied Physics. 2015;48(40):405501.

[51]     Xu Z, Zheng Q-S, Chen G. Elementary building blocks of graphene-nanoribbon-based electronic devices. Applied Physics Letters. 2007;90(22):223115.

[52]     Zhong H, Xu K, Liu Z, Xu G, Shi L, Fan Y, et al. Charge transport mechanisms of graphene/semiconductor Schottky barriers: A theoretical and experimental study. Journal of Applied Physics. 2014;115(1):013701.

**Table 1.** Tight-binding parameters for phosphorene, graphene, and the interface regions. The hopping parameters of phosphorene region are depicted in figure 1. The hopping parameter of graphene region is taken up to third-nearest-neighbor, where $t_1$, $t_2$, and $t_3$ denote the nearest-, second-nearest-, and third-nearest-neighbor hoppings, respectively (all units are in eV).

| Phosphorene | | | | | | Graphene | | | | Interface |
|---|---|---|---|---|---|---|---|---|---|---|
| $\varepsilon_P$ | $t_1$ | $t_2$ | $t_3$ | $t_4$ | $t_5$ | $\varepsilon_C$ | $t_1$ | $t_2$ | $t_3$ | $t_{C-P}$ |
| 0.40 | -1.220 | 3.665 | -0.205 | -0.105 | -0.055 | -0.45 | -2.70 | -0.20 | -0.18 | -1.656 |

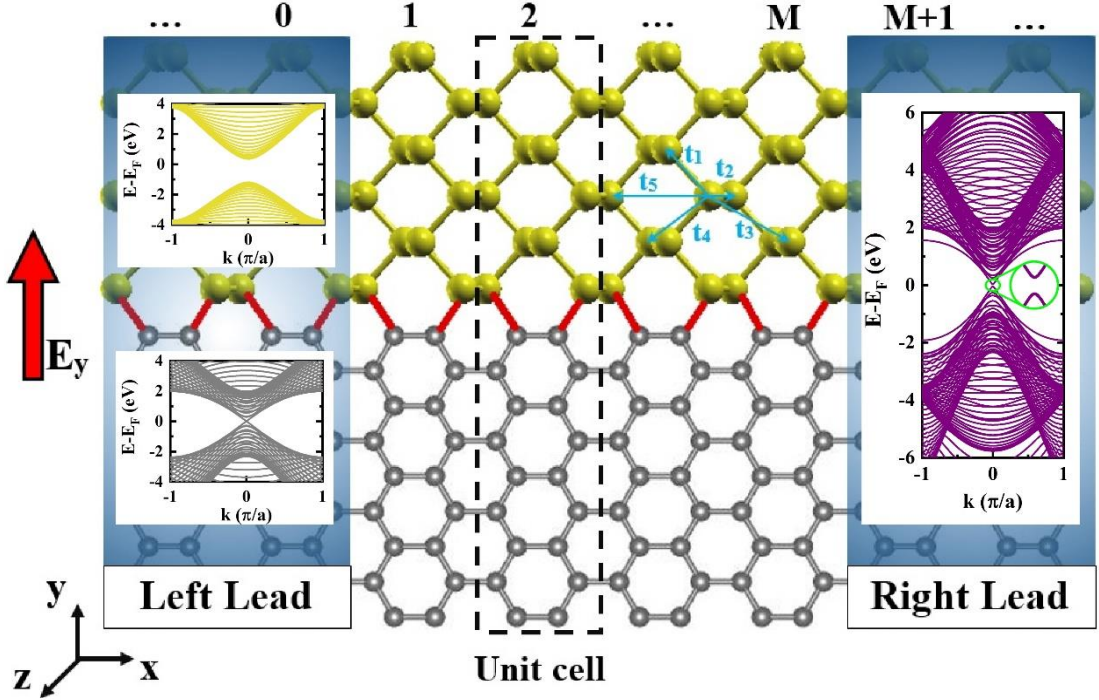

Figure 1. The schematic model of an armchair edged phosphorene/graphene heterostructure attached to two semi-infinite leads. The central region contains M unit cells. The upper left panel: the band structure of 30-APNR,

indicating the semiconducting feature. The lower left panel: the band structure of 41-AGNR showing the metallic feature. The right panel: the band structure of 30-APNR/41-AGNR heterostructure, indicating a small gap of 0.08 eV.

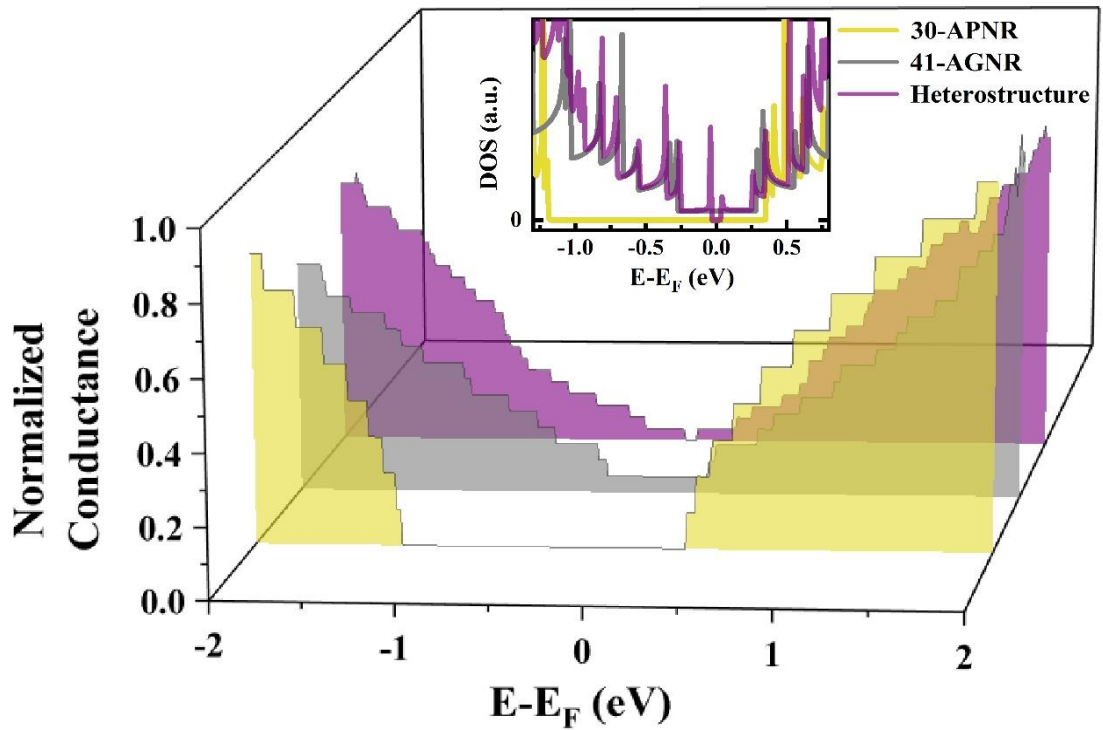

Figure 2. The conductance spectra of 30-APNR/41-AGNR heterostructure (purple region), and the corresponding isolated ribbons: 30-APNR (yellow region) and 41-AGNR (gray region). The inset shows the corresponding density of states spectra. The APNR as a semiconductor shows an energy gap of 1.55 eV, the AGNR shows the metallic feature, and the heterostructure has a small gap of 0.08 eV.

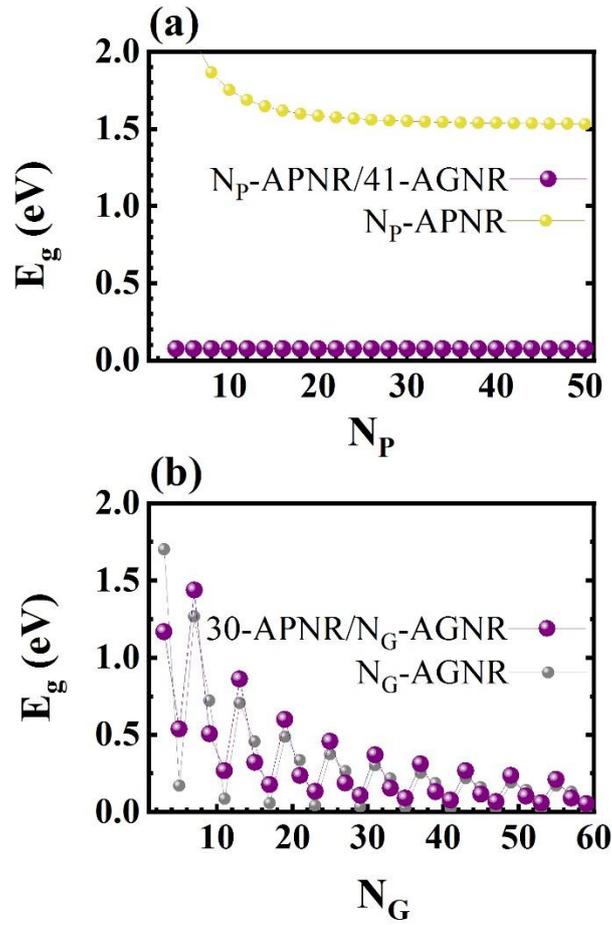

Figure 3. The bandgap of (a) APNR and $N_P$-APNR/41-AGNR heterostructure as functions of $N_P$, and (b) AGNR and 30-APNR/$N_G$-AGNR heterostructure as functions of $N_G$. The $N_P$-APNR/41-AGNR heterostructure is a narrow-gaped system with the semiconductor/metal interface, in which the hole transfer occurs from APNR into 41-AGNR.

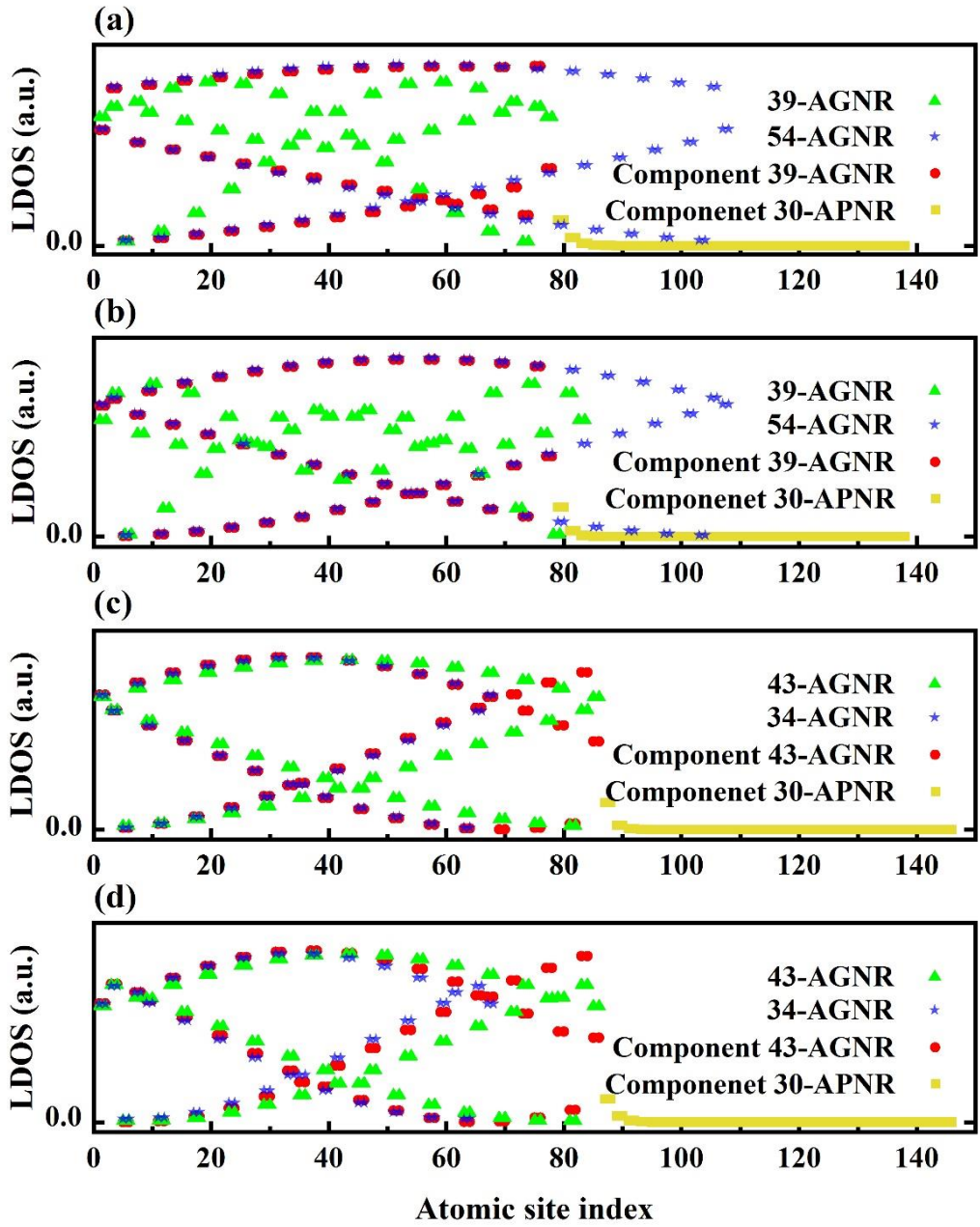

Figure 4. The LDOS of (a) CBM and (b) VBM of 30-APNR/39-AGNR heterostructure, isolated 39-AGNR, and isolated 54-AGNR. LDOS of the component 39-AGNR mostly matches with isolated 54-AGNR rather than isolated 39-AGNR. Also, LDOS of (c) CBM and (d) VBM of 30-APNR/43-AGNR heterostructure, isolated 43-AGNR, and isolated 34-AGNR. LDOS of the component 43-AGNR mostly matches isolated 34-AGNR rather than isolated 43-AGNR.

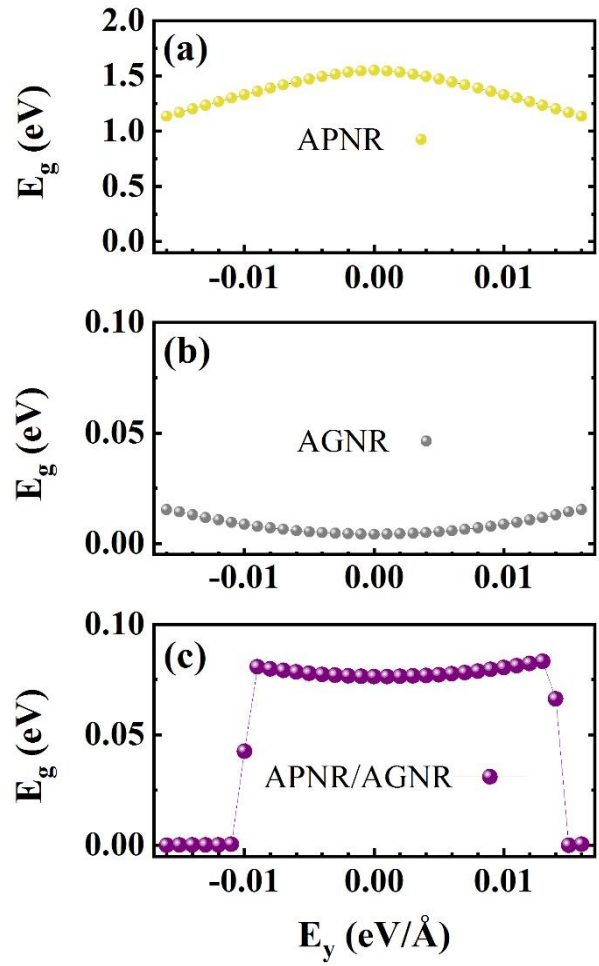

Figure 5. The bandgap of (a) 30-APNR, (b) 41-AGNR, and (c) 30-APNR/41-AGNR heterostructure as a function of the transverse electric field. The semiconductor-metal phase transition occurs by tuning the transverse electric field.

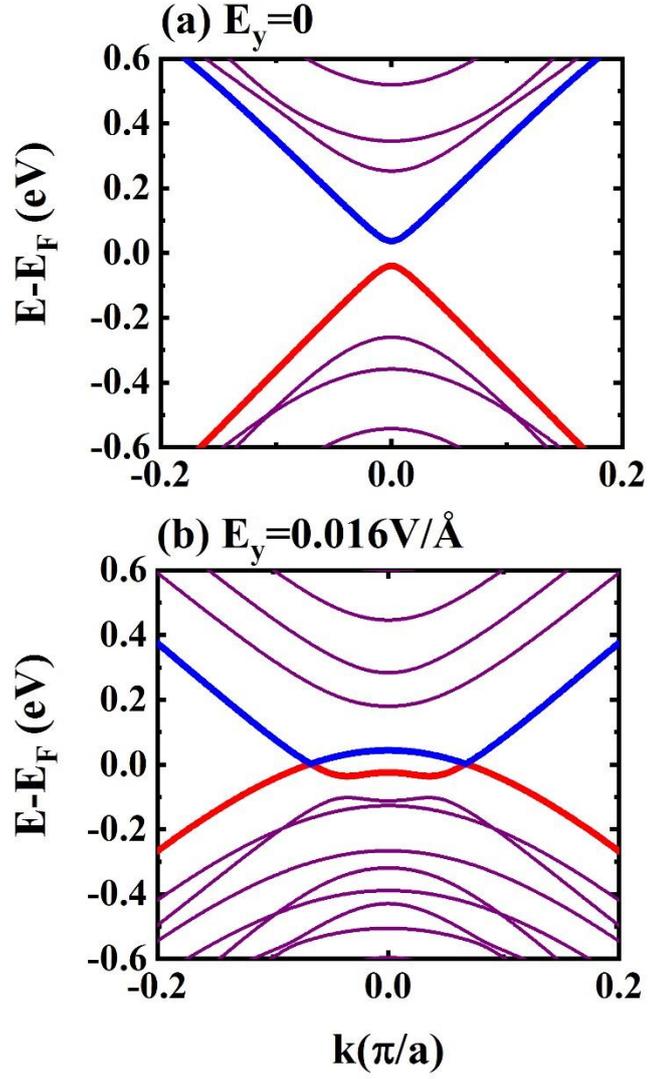

Figure 6. The band structure of 30-APNR/41-AGNR heterostructure when the applied transverse electric field is set to (a) zero and (b) 0.016 eV. The edge states are colored in blue and red. In the absence of the electric field, the edge states are gapped, implying that the system behaves as a band insulator, while under the transverse electric field of $E_y = 0.016$ V/Å the edge states are gapless, implying that the system is a topological insulator.